
 \PHYSREV
\overfullrule0pt

 \newtoks\slashfraction
 \slashfraction={.13}
 \def\slash#1{\setbox0\hbox{$ #1 $}
 \setbox0\hbox to \the\slashfraction\wd0{\hss \box0}/\box0 }

  \def\Buildrel#1\under#2{\mathrel{\mathop{#2}\limits_{#1}}}

\def\lozenge{\boxit{\hbox to 1.5pt{%
             \vrule height 1pt width 0pt \hfill}}}

\def\L{{\cal L}}
\def\M{{\cal M}}
\def\H{{\cal H}}

 \doublespace
 \pubnum{59xx}
\date{October  1992}
 \pubtype{T/E}
 \titlepage
\vfill
 \title{Zero Modes in a $c = 2$ Matrix Model
 \doeack}
 \author{Ovid C. Jacob
 }
 \SLAC

\abstract
Recently
\REF\dk{Simon Dalley and Igor Klebanov,'Light Cone Quantization
of the $c=2$ Matrix Model', PUPT-1333, hepth@xxx/920705}
\refend
Dalley and Klebanov proposed a light-cone quantized study of
the $c=2$ matrix model, but which ignores $k^{+}=0$ contributions.
Since the non-critical string limit of the matrix model involves
taking the parameters $\lambda$ and $\mu$ of the matrix model
to a critical point, zero modes of the field might be important in this study.
The constrained light-cone quantization (CLCQ) approach of
Heinzl, Krusche and Werner is  applied .
It is found that there is coupling between the zero mode sector and
the rest of the theory, hence CLCQ should be implemented.

\vfill
 \endpage

\chapter{ Introduction}

Light-cone quantization has proven to be a very fruitful approach
to study   quantum field theories in 1+1-dimensions
\REF\bp{Hans-Christian Pauli and Stanley Brodsky, Phys. Rev. \us{D32}
1993, (1985); Phys. Rev. \us{D32}, 2001, (1985); Thomas Eller,
Hans-Christian Pauli and Stanley Brodsky, Phys. Rev. \us{D35}, 1493,
(1986); F. Lenz, M. Thies, S. Levit and K.Yazaki, Ann. Phys. \us{208},
1,(1991) }
\refend ,
(as well as for studying non-critical strings \refmark\dk,)
since it is possible to define a sequence of compact
operators of increased 'exactness' by increasing the a 'resolving'
parameter K which governs the total momentum of the system.

It is known that the light- cone
vacuum is trivial, i.e. it is identical to the
perturbative vacuum , and it was not understood how one would get
non-trivial topological behaviour like symmetry breaking.
 Only recently have we started to understand how to do
this
\REF\ntrivac{
   Heinzl, Krusche and Werner like in Phys. Lett. B, \us{256},55 (1991)
or Z. Phys. A, \us{334}, 443 (1989) as well as Regensburg preprints
TPR 91-23, TPR-91-20 and TPR-90-43, TPR-90-44;
Gary McCartor and David Robertson, Zeit. f. Phys. \us{C53}, 679, (1992);
see also K. Hornbostel's preprint   CLNS 91/1078 (1991) as well as
G. McCartor, SMU preprint SMUHEP/91-02 }
\refend  .

This can be seen  by noting that light-cone quantization
is quantization of a system with constraints
\REF\dirac{ P.A.M. Dirac, Can. J. Math.  \us{1},1 (1950).}
\refend
\REF\hrt{Andrew Hanson, Tullio Regge, Claudio Teitelboim, Constrained
Hamiltonian systems, Academia Nazionale dei Lincei, Rome, 1976}
\refend  , as   the Regensburg group has  noted.

Essentially a classical system can have constraints if there are
some q's which don't have conjugate p's. If we want to pass to
the quantum system, we cannot apply the canonical quantization
procedure.
In such cases we need to
examine the nature of these constraints. Dirac calls these constraints
first class if they commute among themselves, and second class if
they do not.

For the first class case we \b{can} use  the
canonical quantization procedure, but need to restrict the Hilbert
space of allowed states upon which the quantum operators are allowed
to operate by imposing these first class constraints on them. Theories
with a gauge symmetry are such an example.

For the second class, Dirac found that by modifying the Poisson
braket he was able to define a consistent set of rules
for quantization. In the case of light-cone quantization, this is the
case we have to deal with.

Lack of implementation of a consistent quantization shows itself in
unphysical aspects in the theory such as breakdown of Lorentz invariance,
lack of gauge invariance or presence of ghosts .
This has also shown up in light-cone quantization. Recently, Burkardt
and Langnau showed that naive light-cone quantization of a Yukawa
model leads to  problems with  rotational invariance
\REF\rotalex{Matthias Burkardt and Alex Langnau, Phys. Rev.
\us{D44}, 1187 (1991); Phys. Rev. \us{D44}, 3859}
\refend  .
Their solution is to add appropriate counterterms in the Hamiltonian
and which restores the symmetry.
\REF\ohiost{Or see
Robert J. Perry, Avaroth Harindranath and Kenneth
G. Wilson, Phys. Rev. Lett. \us{65}, 2959 (1990); A. Harindranath and
Robert J. Perry, Phys. Rev. \us{D43}, 4921 (1991) for an approach
similar in spirit, though quite different in philosophy and method }
\refend  .

Recently, Dalley and Klebanov \refmark\dk
applied DLCQ to the study of $c=2$
matrix model. In this approach, one takes a double scaling limit
\REF\GM{D.J. Gross and A.A. Migdal, Phys. Rev. Lett. \us{64}, 717
(1990); M. Douglas and S. Shenker, Nucl. Phys. \us{B335}, 635(1990)}
\refend
of a matrix model
\REF\BIPZ{E. Brezin, C. Itzykson, G. Parisi and J.-B. Zuber,
Comm. Math. Phys. \us{59}, 35 (1978)}
\refend
as a way to study $c=2$ non-critical strings
\REF\Thorn{C. B. Thorn, Phys. Lett. \us{242}, 364 (1990)}
\refend .
 In this limit, one lets the coupling $g$ go to 1 and N go to $\infty$
but one keeps fixed some product of N and $(g-1)$.
In their work, Dalley and Klebanov took the N going to $\infty$ limit
first and then looked for critical coupling.
They   interpreted the appearance of continuous
states in the mass spectrum as the appearance of the Liouville
mode
\REF\CT{T. L. Curtright and C. B. Thorn, Phys. Rev. Lett \us{48}
, 1309(1982) }
\refend  .
They also found the presence of tachyonic states, which they identified
with those of bosonic strings.
In this paper I will to study the effect  zero modes of the field might have
on this $c=2$   matrix model. The outline of the paper is as follows:
first I will study a  similar model , the $\phi^{3}$ model , in
CLCQ. Then I will apply the results obtained to the model of
Dalley and Klebanov.

\chapter{ Constrained Light-Cone Quantization of the $\phi^{3}$ Model}
It is assumed that the reader is familiar with the way one implements
CLCQ  from previous work \refmark\dirac ,\refmark\ntrivac, so it
 In this study I will follow
 the work of Werner and collaborators \refmark\ntrivac.

Consider then the  Langrangian
$$ \L_{C}=\L_{0}+\L_{I} $$
where
$$ \L_{0}= {1 \over 2} \partial_{\mu} \phi \partial^{\mu} \phi
+ {1\over2}\mu \phi^{2} $$
and
$$ \L_{I}= - {\lambda \phi^{3} \over 3! } $$
In light-cone coordinates this is
$$ \L_{C}= \partial_{+} \phi \partial_{-} \phi
+{1\over 2}\mu  \phi^{2} -{\lambda \over 3!}\phi^{3} $$

To solve the theory we introduce a box of dimensions 2L in the
$x^{-}$ directions.
Define now the zero mode background   field $\omega $
$$ \omega(x^{+}) =  P*\phi(x^{+},x^{-}) =
{1\over 2L} \int_{-L}^{L} dx^{-} \phi(x^{+},x^{-})$$
This is the P-space or zero mode
projection of the scalar field $\phi$. Its complement
Q is defined thus
$$ \varphi(x^{+},x^{-})=Q*\phi (x^{+},x^{-})=\phi (x^{+},x^{-})-
P*\phi (x^{+},x^{-}) $$
The free part of the Lagrangian becomes :
$$ \L_{0} = \partial_{+}\varphi \partial_{-}\varphi -{1\over2}
\mu (\varphi + \omega)^{2} $$
Then the interaction part of the Lagrangian becomes
$$\L_{I}={1\over2}\mu (\varphi + \omega)^{2} -{\lambda \over 3!}
(\varphi +   \omega)^{3} $$
$$     ={1\over2}\mu\varphi^{2} +{1\over2}\omega^{2} +\mu\varphi\omega -
{\lambda \over 3!}(\varphi^{3}+\omega^{3}+3 \varphi^{2}\omega +
3 \varphi \omega^{2} ) $$
The canonical $\varphi$ momentum is
$$ \pi_{\varphi}={\partial\L_{C} \over \partial (\partial_{+} \varphi)}
=\partial_{-} \varphi $$
leading to the constraint
$$ \theta_{1}(x^{+},x^{-}) = \pi_{\varphi}(x^{+},x^{-}) - \partial
_{-}\varphi(x^{+},x^{-}) $$
The result for the background field $\omega$ is
$$ \pi_{\omega} \approx 0$$
and  the next constraint is
$$ \theta_{2}(x^{+}) = \pi_{\omega}(x^{+}) $$
These degrees of freedom have the following canonical commutation
$$\big\{ \varphi(x^{+},x^{-}),\pi_{\varphi}(x^{+},y^{-}) \big\} =
\delta (x^{-} - y^{-}) $$
and
$$\big\{ \omega , \pi_{\omega} \big\} = 1$$
We obtain the following result for the canonical Hamiltonian $\H_{C}$
$$\H_{C}={1\over2}\mu (\varphi+\omega)^{2} -{\lambda \over 3!}(\varphi +
 \omega)^{3} $$
We implement the   constraints by adding extra terms to the
canonical Hamiltonian $H_{C}$
to obtain the 'proper' Hamiltonian $H_{p}$
$$ H_{P}(x^{+}) = H_{C}(x^{+}) +
\int_{-L}^{L}dx^{-} u_{1}(x^{+},x^{-})\theta_{1}(x^{+},x^{-})+
\theta_{2}(x^{+})u_{2}(x^{+})2 L  $$
The Lagrange multipliers $u_{1}$ and $u_{2}$ are determined by
requiring that they have zero Poisson commutator with the modified
Hamiltonian $H_{P}$ . If the first commutator is not zero, we continue
to take commutators until  zero \us{is} obtained \refmark\hrt .
We use the cannonical commutation expressions for $\varphi$ and
$\omega $ introduced above.
The following expression  is  obtained    for $u_{1}$
$$u_{1}(x^{+},x^{-}) = \int_{-L}^{+L}dy^{-} G^{Q}(x^{-},y^{-}){1\over4}
\big\{\mu\varphi(x^{+},y^{-})+\mu\omega(x^{+})+$$
$$ +{\lambda \over 2!}(\varphi(x^{+},y^{-})^2+
 2\varphi(x^{+},y^{-})\omega(x^{+}) +  \omega(x^{+})^{2} ) \big\} $$
where $G^{Q}(x,y)$ is the Q-projected Green's function
$$G^{Q}(x,y)={1\over2} sgn(x-y)-{x-y \over 2L} $$
To determine  $u_{2}$ we need to work a bit harder.
The first calculation   gives
$$\partial_{+}\theta_{2}=\big\{\theta_{2},\H_{P} \big\} $$
$$\partial_{+}\theta_{2}=-\mu \omega(x^{+}) +{\lambda \over 2!}
\omega(x^{+})^{2}+$$
$$+\int_{-L}^{L}dx^{-}{\lambda \over 2!}\big\{
\varphi(x^{+},x^{-})^{2} +2\omega(x^{+})\varphi(x^{+},x^{-}) \big\}
 = \theta_{3}(x^{+})\approx 0$$
Since this equation  does not give us an expression for $u_2$, we
take the commutator of the new constraint, $\theta_3$, with the
full Hamiltonian. We get the following results for the commutators with
$\theta_1$ and $\theta_2$ respectively:
$$\big\{\theta_3, \theta_1 \big\} = {1 \over 2L}(\lambda \varphi(x^{+},
x^{-})-\mu ) $$
and
$$\big\{\theta_3, \theta_2 \big\} = {1 \over 2L}(2 \lambda \omega(x^{+})
-\mu ) $$
Putting this in the commutator with $\H_P$, we get
$$u_{2}(x^{+})={-1 \over 2L}\int_{-L}^{L} dx^{-}{ \lambda \varphi(x^{-},
x^{+}) - \mu  \over 2 \lambda \omega(x^{+}) - \mu}u_{1}(x^{-},x^{+}) $$

To determine the new commutation relations,   construct the Dirac
braket $\{,\}^{*}$
$$ \{A,B\}^{*} = \{A,B\}-\sum_{ij} \{A,\psi_{i}\} \{\psi_{i},
\psi_{j}\}^{-1}\{\psi_{j},B\} $$
and here the $\psi_{i}$'s are all second class constraints, so that
the inverse is meaningful.

With this,  the following Dirac brackets are obtained :
$$ \{\varphi(x^{+},x^{-}),\varphi(x^{+},y^{-}) \}^{*} =
-{1\over4} G^{Q}(x^{-},y^{-}) $$
and
$$ \{\varphi(x^{+},x^{-}),\pi_{\varphi}(x^{+},y^{-}) \}^{*} =
+\partial_{-}^{y^{-}}G^{Q}(x^{-},y^{-}) $$
These  are as expected. The interesting result is that we also get
$$ \{\omega(x^{+}),\varphi(x^{+},x^{-}) \}^{*}=-{1\over4}
\int_{-L}^{L}dy^{-}
G^{Q}(x^{-},y^{-}){\lambda \varphi(x^{+},y^{-})-\mu \over  2 \lambda
\omega(x^{+}) - \mu} $$
This indicates that there is coupling between the  non-zero modes of the field
of
the scalar field and the background field $\omega$, the zero mode of the
scalar field.

There  is also coupling between the zero modes of the field and the momentum of
the
non-zero modes of the field :
$$ \{\omega(x^{+}),\pi_{\varphi}(x^{+},x^{-}) \}^{*}={1\over4}
\int_{-L}^{L}dy^{-}
G^{Q}(x^{-},y^{-}){\lambda \varphi(x^{+},y^{-})-\mu \over ( 2 \lambda
\omega(x^{+}) - \mu)^{2}} 2 \lambda $$
On the other hand we get the following Dirac brackets for the
zero mode of the field
$$ \{\omega(x^{+}),\pi_{\omega}(x^{+}) \}^{*}=
   \{\omega(x^{+}),\omega(x^{+}) \}^{*}=
   \{\pi_{\omega}(x^{+}),\pi_{\omega}(x^{+}) \}^{*}=0 $$
This means that upon Dirac quantization this quantity is not dynamical.
Nonetheless, since it depends on the quantized field $\phi$, and
since it couples to the non-zero part of the field, it is important
in studying non-trivial topological properties of the theory.
\chapter{ The c=2 Matrix Model  }
Let us study now this Dirac (constrained) quantization applied to
the $c=2$ matrix model introduced by Dalley and Klebanov \refmark\dk.
The Lagrangian is
$$\L = Tr\Big\{1/2(\partial_{\alpha}M)^{2}+1/2\mu M^{2}-{\lambda \over
3\sqrt{N} }M^{3} \Big\} $$
where $M(x^{-},x^{+})$ are $N\times N$ hermitian matrices.
We apply now the method of  constrained quantization described above
and get the following for the full Hamiltonian
$$P^{-}(x^{+}) = \int_{-L}^{L} dx^{-}Tr\Big\{{1\over2}\mu M^{2}-{\lambda
\over 3\sqrt{N} } M^{3} \Big\}+ $$
$$ +\int_{-L}^{L}dx^{-}Tr \Big\{U_{1}(x^{+},x^{-})\Theta_{1}(x^{+},x^{-})
\Big\} +Tr\Big\{\Theta_{2}(x^{+})U_{2}(x^{+})2L\Big\} $$
where the $U's$ are the new Lagrange multipliers and the $\Theta's$
the new constraints :
$$ \Theta_{1}(x^{+},x^{-}) =
\Pi_{\M}(x^{-},x^{+})-\partial_{-}\M(x^{-},x^{+}) \approx 0$$
and
$$\Theta_{2}(x^{+}) =\Pi_{\Omega}(x^{+}) \approx 0  $$

The analysis goes through as in the previous chapter - except that
now I have the extra indices, since $TrM^{2}$ means $\sum_{ij}M_{ij}
M_{ji}$; I'll suppress these indices from now on.
As before, I split the field $M$ into  a zero mode part
$$ \Omega(x^{+}) =  P*M(x^{+},x^{-}) =
{1\over 2L} \int_{-L}^{L} dx^{-} M(x^{+},x^{-})$$
and
$$ \M(x^{+},x^{-})=Q*M (x^{+},x^{-})=\delta *M(x^{+},x^{-})-
P*M(x^{+},x^{-}) =M(x^{+},x^{-})-\Omega(x^{+})$$
is the non-zero part. I get the following form for the new hamiltonian
$$\H_{P}=P^{-}(x^{+})=
 \int_{-L}^{L} dx^{-}Tr\Big\{{1\over2}\mu M^{2}-{\lambda
\over 3\sqrt{N} } M^{3} \Big\}+ $$
$$ +\int_{-L}^{L}dx^{-}Tr \Big\{
(\Pi_{\M}(x^{-},x^{+})-\partial_{-}\M(x^{-},x^{+}) )
U_{1}(x^{+},x^{-})\Big\} +
Tr\Big\{\Pi_{\Omega}(x^{+})U_{2}(x^{+})2L\Big\} $$
where the $\Pi_{\M}$ is the momentum canonical to $\M$ and $\Pi_{\Omega}$
the momentum canonical to $\Omega$.
The $U's$ are found to be
$$U_{1}(x^{+},x^{-}) = \int_{-L}^{+L}dy^{-} G^{Q}(x^{-},y^{-}){1\over4}
\Big\{\mu\M(x^{+},y^{-})+\mu\Omega(x^{+})+$$
$$+{\lambda \over 2!}(\M(x^{+},y^{-})^2
+ 2\M(x^{+},y^{-})\Omega(x^{+}) + \Omega(x^{+})^{2} ) \Big\} $$
and
$$U_{2}(x^{+})={-1 \over 2L}\int_{-L}^{L} dx^{-}
(2 \lambda \Omega(x^{+}) - \mu I)^{-1}
( \lambda \M(x^{-},x^{+}) - \mu I) U_{1}(x^{-},x^{+}) $$
As before,
the interesting part is that there is coupling between the zero mode
sector and the non-zero mode sector due to the following commutator
$$ \{\Omega(x^{+}),\M(x^{+},x^{-}) \}^{*}=-{1\over4}
\int_{-L}^{L}dy^{-}
(2 \lambda \Omega(x^{+}) - \mu I) ^{-1}
G^{Q}(x^{-},y^{-})(\lambda \M(x^{+},y^{-}) - \mu I) $$

In this case
there  is also coupling between the zero modes of the field and the momentum of
the
non-zero modes of the field :
$$ \{\Omega(x^{+}),\Pi_{\M}(x^{+},x^{-}) \}^{*}={1\over4}
\int_{-L}^{L}dy^{-}
( 2 \lambda \Omega(x^{+}) - \mu I)^{-2}
G^{Q}(x^{-},y^{-})
(\lambda \M(x^{+},y^{-})-\mu I) 2 \lambda $$
\chapter{Conclusions}
The constrained light-cone quantization indicates that there is
coupling between the zero mode sector and the nonzero mode sector.
This means that the analysis of Dalley and Klebanov \refmark\dk
needs to be redone in light of this result.  It is unclear at this point
if there is still the kind of excitations which Dalley and Klebanov
associated with Liouville mode.
This is because the type of critical behaviour studied might involve
excitations of the zero modes of the field which were previously left out. A
more
careful study is necessary to discover what happens now in the
 double scaling limit of this matrix  model.

There is also the possibility that the tachyonic mode which they discover
might be due to instabilities in the $\phi^{3}$ theory rather than
due to the bosonic string. In a recent paper, Hiller and Swenson
\REF\hs{ John Swenson  and John Hiller, 'Numerical signatures of
vacuum instability in a one-dimensional Wick Cutkosky model on the
light-cone,' Sept. 1992}
studied the Wick-Cutkosky model which is similar to the $\phi^{3}$
model considered by Dalley and Klebanov, and found instabilities in
vacuum , as expected for a cubic theory.
\chapter{Acknowledgements}
I would like to thank Prof. Stanley Brodsky for suggesting this problem
and to Prof. Blankenbecler for his continuing support .
I would also like to thank Igor Klebanov and Dave Robertson   for
useful discussions.
\endpage
\refout
\end
\REF\clcqschw{T. Heinzl, S. Krusche and E. Werner, 'The Fermionic
Schwinger Model in Light Cone  Quantization,' Phys. Lett. B, \us{275},
410, (1992) }
\refend
\REF\clcqschwa{T. Heinzl, S. Krusche and E. Werner, 'Non-Trivial Vacuum
Structure in Light Cone  Quantum Field Theory,' Nucl. Phys.  ,
 A, \us{532},429c, (1991) }
\refend
\REF\clcqschwb{T. Heinzl, S. Krusche and E. Werner, 'Non-Perturbative
Vacua in Light Cone  Quantum Field Threory,' Nucl. Phys. (Proc.
Suppl.), B, \us{23},182, (1991) }
\refend
\REF\clcqphi{T. Heinzl, S. Krusche and E. Werner,
'Light Cone  Quantization of Scalar Field Theories,' TPR-91-20 }
\refend
\REF\phitreeplus{T. Heinzl, S. Krusche and E. Werner, 'Non-Perturbative
 Light Cone  Quantum  Field Theories Beyond the Tree Level,' TPR-92-16 }
\refend
\REF\phitreezero{T. Heinzl, S. Krusche and E. Werner, 'Zero Mode
 Corrections in Light-Cone  Quantum  Field Theory ,' TPR-92-17 }
\refend
\REF\phitree{
T. Heinzl, S. Krusche and E. Werner, 'Spontaneous Symmetry Breaking
 in Light Cone  Quantum Field Theory' Phys. Lett. B, \us{272},
 54, (1991) }
\refend
\REF\ntrivac{
T. Heinzl, S. Krusche and E. Werner like in Phys. Lett. B,
\us{256},55 (1991)
or Z. Phys. A, \us{334}, 443 (1989); see also K. Hornbostel's preprint
CLNS 91/1078 (1991);Gary McCartor, Z. Phys. \us{C41}, 271 (1988)}
\refend
\REF\dirac{ P.A.M. Dirac, Rev. Mod. Phys.\us{21},392 (1949).}
\refend
\REF\alexmat{Matthias Burkardt and Alex Langnau, Phys. Rev.
\us{D44}, 3859 (1991)}
\refend
\REF\ohiost{Robert J. Perry, Avaroth Harindranath and Kenneth
G. Wilson, Phys. Rev. Lett. \us{65}, 2959 (1990); A. Harindranath and
Robert J. Perry, Phys. Rev. \us{D43}, 4921 (1991)}
\refend
\REF\bsdisc{Ovid C. Jacob, 'Discrete Symmetries for the Bound State
Problem of Positronium in Light-Cone Quantization,' SLAC-PUB-58xx,
 June 1992 .}
\refend
\REF\dba{P. A. M. Dirac, Can. Jour. Math., \us{1}, 1, (1950); 'Lectures
in Quantum Mechanics', Benjamin, NY, 1964;
P. G. Bergmann, Helv. Phys. Acta (Suppl.), \us{4}, 79 (1956) }
\refend .
\REF\zeropert{Gary McCartor and David Robertson, 'Bosonic Zero
Modes in DLCQ,' SMUHEP/91-04 Oct.   1991}
\refend
\end
\refout
\end
==:B:==
================================================================================
Received: by SLACVM (Mailer R2.08 R208004) id 0129;
          Fri, 28 May 93 15:50:03 PST
Date: Fri, 28 May 93 15:50:02 PST
{}From: Network Mailer <MAILER@SLACVM>
Subject: Undelivered mail
To: JACOB@SLACVM

Your mail was not delivered to some or all of its
intended recipients for the following reason(s):

Message syntax is unrecognizable.

--------------------RETURNED MAIL FILE--------------------
Date: Fri, 28 May 1993   15:44 -0800 (PST)
{}From: JACOB@SLACVM
To:   hep-th@xxx.lanl.gov
Subject: put
\\
 Zero Modes in a $c = 2$ Matrix Model , by  Ovid C. Jacob , phyzxx, 14
 pages, SLAC-PUB-59xx .
\\
Recently
\REF\dk{Simon Dalley and Igor Klebanov,'Light Cone Quantization
of the $c=2$ Matrix Model', PUPT-1333, hepth@xxx/920705}
\refend
Dalley and Klebanov proposed a light-cone quantized study of
the $c=2$ matrix model, but which ignores $k^{+}=0$ contributions.
Since the non-critical string limit of the matrix model involves
taking the parameters $\lambda$ and $\mu$ of the matrix model
to a critical point, zero modes of the field might be important in this study.
The constrained light-cone quantization (CLCQ) approach of
Heinzl, Krusche and Werner is  applied .
It is found that there is coupling between the zero mode sector and
the rest of the theory, hence CLCQ should be implemented.

\\
 \PHYSREV
\overfullrule0pt

 \newtoks\slashfraction
 \slashfraction={.13}
 \def\slash#1{\setbox0\hbox{$ #1 $}
 \setbox0\hbox to \the\slashfraction\wd0{\hss \box0}/\box0 }

  \def\Buildrel#1\under#2{\mathrel{\mathop{#2}\limits_{#1}}}

\def\lozenge{\boxit{\hbox to 1.5pt{%
             \vrule height 1pt width 0pt \hfill}}}

\def\L{{\cal L}}
\def\M{{\cal M}}
\def\H{{\cal H}}

 \doublespace
 \pubnum{59xx}
\date{October  1992}
 \pubtype{T/E}
 \titlepage
\vfill
 \title{Zero Modes in a $c = 2$ Matrix Model
 \doeack}
 \author{Ovid C. Jacob
 }
 \SLAC

\abstract
Recently
\REF\dk{Simon Dalley and Igor Klebanov,'Light Cone Quantization
of the $c=2$ Matrix Model', PUPT-1333, hepth@xxx/920705}
\refend
Dalley and Klebanov proposed a light-cone quantized study of
the $c=2$ matrix model, but which ignores $k^{+}=0$ contributions.
Since the non-critical string limit of the matrix model involves
taking the parameters $\lambda$ and $\mu$ of the matrix model
to a critical point, zero modes of the field might be important in this study.
The constrained light-cone quantization (CLCQ) approach of
Heinzl, Krusche and Werner is  applied .
It is found that there is coupling between the zero mode sector and
the rest of the theory, hence CLCQ should be implemented.

\vfill
 \endpage

\chapter{ Introduction}

Light-cone quantization has proven to be a very fruitful approach
to study   quantum field theories in 1+1-dimensions
\REF\bp{Hans-Christian Pauli and Stanley Brodsky, Phys. Rev. \us{D32}
1993, (1985); Phys. Rev. \us{D32}, 2001, (1985); Thomas Eller,
Hans-Christian Pauli and Stanley Brodsky, Phys. Rev. \us{D35}, 1493,
(1986); F. Lenz, M. Thies, S. Levit and K.Yazaki, Ann. Phys. \us{208},
1,(1991) }
\refend ,
(as well as for studying non-critical strings \refmark\dk,)
since it is possible to define a sequence of compact
operators of increased 'exactness' by increasing the a 'resolving'
parameter K which governs the total momentum of the system.

It is known that the light- cone
vacuum is trivial, i.e. it is identical to the
perturbative vacuum , and it was not understood how one would get
non-trivial topological behaviour like symmetry breaking.
 Only recently have we started to understand how to do
this
\REF\ntrivac{
   Heinzl, Krusche and Werner like in Phys. Lett. B, \us{256},55 (1991)
or Z. Phys. A, \us{334}, 443 (1989) as well as Regensburg preprints
TPR 91-23, TPR-91-20 and TPR-90-43, TPR-90-44;
Gary McCartor and David Robertson, Zeit. f. Phys. \us{C53}, 679, (1992);
see also K. Hornbostel's preprint   CLNS 91/1078 (1991) as well as
G. McCartor, SMU preprint SMUHEP/91-02 }
\refend  .

This can be seen  by noting that light-cone quantization
is quantization of a system with constraints
\REF\dirac{ P.A.M. Dirac, Can. J. Math.  \us{1},1 (1950).}
\refend
\REF\hrt{Andrew Hanson, Tullio Regge, Claudio Teitelboim, Constrained
Hamiltonian systems, Academia Nazionale dei Lincei, Rome, 1976}
\refend  , as   the Regensburg group has  noted.

Essentially a classical system can have constraints if there are
some q's which don't have conjugate p's. If we want to pass to
the quantum system, we cannot apply the canonical quantization
procedure.
In such cases we need to
examine the nature of these constraints. Dirac calls these constraints
first class if they commute among themselves, and second class if
they do not.

For the first class case we \b{can} use  the
canonical quantization procedure, but need to restrict the Hilbert
space of allowed states upon which the quantum operators are allowed
to operate by imposing these first class constraints on them. Theories
with a gauge symmetry are such an example.

For the second class, Dirac found that by modifying the Poisson
braket he was able to define a consistent set of rules
for quantization. In the case of light-cone quantization, this is the
case we have to deal with.

Lack of implementation of a consistent quantization shows itself in
unphysical aspects in the theory such as breakdown of Lorentz invariance,
lack of gauge invariance or presence of ghosts .
This has also shown up in light-cone quantization. Recently, Burkardt
and Langnau showed that naive light-cone quantization of a Yukawa
model leads to  problems with  rotational invariance
\REF\rotalex{Matthias Burkardt and Alex Langnau, Phys. Rev.
\us{D44}, 1187 (1991); Phys. Rev. \us{D44}, 3859}
\refend  .
Their solution is to add appropriate counterterms in the Hamiltonian
and which restores the symmetry.
\REF\ohiost{Or see
Robert J. Perry, Avaroth Harindranath and Kenneth
G. Wilson, Phys. Rev. Lett. \us{65}, 2959 (1990); A. Harindranath and
Robert J. Perry, Phys. Rev. \us{D43}, 4921 (1991) for an approach
similar in spirit, though quite different in philosophy and method }
\refend  .

Recently, Dalley and Klebanov \refmark\dk
applied DLCQ to the study of $c=2$
matrix model. In this approach, one takes a double scaling limit
\REF\GM{D.J. Gross and A.A. Migdal, Phys. Rev. Lett. \us{64}, 717
(1990); M. Douglas and S. Shenker, Nucl. Phys. \us{B335}, 635(1990)}
\refend
of a matrix model
\REF\BIPZ{E. Brezin, C. Itzykson, G. Parisi and J.-B. Zuber,
Comm. Math. Phys. \us{59}, 35 (1978)}
\refend
as a way to study $c=2$ non-critical strings
\REF\Thorn{C. B. Thorn, Phys. Lett. \us{242}, 364 (1990)}
\refend .
 In this limit, one lets the coupling $g$ go to 1 and N go to $\infty$
but one keeps fixed some product of N and $(g-1)$.
In their work, Dalley and Klebanov took the N going to $\infty$ limit
first and then looked for critical coupling.
They   interpreted the appearance of continuous
states in the mass spectrum as the appearance of the Liouville
mode
\REF\CT{T. L. Curtright and C. B. Thorn, Phys. Rev. Lett \us{48}
, 1309(1982) }
\refend  .
They also found the presence of tachyonic states, which they identified
with those of bosonic strings.
In this paper I will to study the effect  zero modes of the field might have
on this $c=2$   matrix model. The outline of the paper is as follows:
first I will study a  similar model , the $\phi^{3}$ model , in
CLCQ. Then I will apply the results obtained to the model of
Dalley and Klebanov.

\chapter{ Constrained Light-Cone Quantization of the $\phi^{3}$ Model}
It is assumed that the reader is familiar with the way one implements
CLCQ  from previous work \refmark\dirac ,\refmark\ntrivac, so it
 In this study I will follow
 the work of Werner and collaborators \refmark\ntrivac.

Consider then the  Langrangian
$$ \L_{C}=\L_{0}+\L_{I} $$
where
$$ \L_{0}= {1 \over 2} \partial_{\mu} \phi \partial^{\mu} \phi
+ {1\over2}\mu \phi^{2} $$
and
$$ \L_{I}= - {\lambda \phi^{3} \over 3! } $$
In light-cone coordinates this is
$$ \L_{C}= \partial_{+} \phi \partial_{-} \phi
+{1\over 2}\mu  \phi^{2} -{\lambda \over 3!}\phi^{3} $$

To solve the theory we introduce a box of dimensions 2L in the
$x^{-}$ directions.
Define now the zero mode background   field $\omega $
$$ \omega(x^{+}) =  P*\phi(x^{+},x^{-}) =
{1\over 2L} \int_{-L}^{L} dx^{-} \phi(x^{+},x^{-})$$
This is the P-space or zero mode
projection of the scalar field $\phi$. Its complement
Q is defined thus
$$ \varphi(x^{+},x^{-})=Q*\phi (x^{+},x^{-})=\phi (x^{+},x^{-})-
P*\phi (x^{+},x^{-}) $$
The free part of the Lagrangian becomes :
$$ \L_{0} = \partial_{+}\varphi \partial_{-}\varphi -{1\over2}
\mu (\varphi + \omega)^{2} $$
Then the interaction part of the Lagrangian becomes
$$\L_{I}={1\over2}\mu (\varphi + \omega)^{2} -{\lambda \over 3!}
(\varphi +   \omega)^{3} $$
$$     ={1\over2}\mu\varphi^{2} +{1\over2}\omega^{2} +\mu\varphi\omega -
{\lambda \over 3!}(\varphi^{3}+\omega^{3}+3 \varphi^{2}\omega +
3 \varphi \omega^{2} ) $$
The canonical $\varphi$ momentum is
$$ \pi_{\varphi}={\partial\L_{C} \over \partial (\partial_{+} \varphi)}
=\partial_{-} \varphi $$
leading to the constraint
$$ \theta_{1}(x^{+},x^{-}) = \pi_{\varphi}(x^{+},x^{-}) - \partial
_{-}\varphi(x^{+},x^{-}) $$
The result for the background field $\omega$ is
$$ \pi_{\omega} \approx 0$$
and  the next constraint is
$$ \theta_{2}(x^{+}) = \pi_{\omega}(x^{+}) $$
These degrees of freedom have the following canonical commutation
$$\big\{ \varphi(x^{+},x^{-}),\pi_{\varphi}(x^{+},y^{-}) \big\} =
\delta (x^{-} - y^{-}) $$
and
$$\big\{ \omega , \pi_{\omega} \big\} = 1$$
We obtain the following result for the canonical Hamiltonian $\H_{C}$
$$\H_{C}={1\over2}\mu (\varphi+\omega)^{2} -{\lambda \over 3!}(\varphi +
 \omega)^{3} $$
We implement the   constraints by adding extra terms to the
canonical Hamiltonian $H_{C}$
to obtain the 'proper' Hamiltonian $H_{p}$
$$ H_{P}(x^{+}) = H_{C}(x^{+}) +
\int_{-L}^{L}dx^{-} u_{1}(x^{+},x^{-})\theta_{1}(x^{+},x^{-})+
\theta_{2}(x^{+})u_{2}(x^{+})2 L  $$
The Lagrange multipliers $u_{1}$ and $u_{2}$ are determined by
requiring that they have zero Poisson commutator with the modified
Hamiltonian $H_{P}$ . If the first commutator is not zero, we continue
to take commutators until  zero \us{is} obtained \refmark\hrt .
We use the cannonical commutation expressions for $\varphi$ and
$\omega $ introduced above.
The following expression  is  obtained    for $u_{1}$
$$u_{1}(x^{+},x^{-}) = \int_{-L}^{+L}dy^{-} G^{Q}(x^{-},y^{-}){1\over4}
\big\{\mu\varphi(x^{+},y^{-})+\mu\omega(x^{+})+$$
$$ +{\lambda \over 2!}(\varphi(x^{+},y^{-})^2+
 2\varphi(x^{+},y^{-})\omega(x^{+}) +  \omega(x^{+})^{2} ) \big\} $$
where $G^{Q}(x,y)$ is the Q-projected Green's function
$$G^{Q}(x,y)={1\over2} sgn(x-y)-{x-y \over 2L} $$
To determine  $u_{2}$ we need to work a bit harder.
The first calculation   gives
$$\partial_{+}\theta_{2}=\big\{\theta_{2},\H_{P} \big\} $$
$$\partial_{+}\theta_{2}=-\mu \omega(x^{+}) +{\lambda \over 2!}
\omega(x^{+})^{2}+$$
$$+\int_{-L}^{L}dx^{-}{\lambda \over 2!}\big\{
\varphi(x^{+},x^{-})^{2} +2\omega(x^{+})\varphi(x^{+},x^{-}) \big\}
 = \theta_{3}(x^{+})\approx 0$$
Since this equation  does not give us an expression for $u_2$, we
take the commutator of the new constraint, $\theta_3$, with the
full Hamiltonian. We get the following results for the commutators with
$\theta_1$ and $\theta_2$ respectively:
$$\big\{\theta_3, \theta_1 \big\} = {1 \over 2L}(\lambda \varphi(x^{+},
x^{-})-\mu ) $$
and
$$\big\{\theta_3, \theta_2 \big\} = {1 \over 2L}(2 \lambda \omega(x^{+})
-\mu ) $$
Putting this in the commutator with $\H_P$, we get
$$u_{2}(x^{+})={-1 \over 2L}\int_{-L}^{L} dx^{-}{ \lambda \varphi(x^{-},
x^{+}) - \mu  \over 2 \lambda \omega(x^{+}) - \mu}u_{1}(x^{-},x^{+}) $$

To determine the new commutation relations,   construct the Dirac
braket $\{,\}^{*}$
$$ \{A,B\}^{*} = \{A,B\}-\sum_{ij} \{A,\psi_{i}\} \{\psi_{i},
\psi_{j}\}^{-1}\{\psi_{j},B\} $$
and here the $\psi_{i}$'s are all second class constraints, so that
the inverse is meaningful.

With this,  the following Dirac brackets are obtained :
$$ \{\varphi(x^{+},x^{-}),\varphi(x^{+},y^{-}) \}^{*} =
-{1\over4} G^{Q}(x^{-},y^{-}) $$
and
$$ \{\varphi(x^{+},x^{-}),\pi_{\varphi}(x^{+},y^{-}) \}^{*} =
+\partial_{-}^{y^{-}}G^{Q}(x^{-},y^{-}) $$
These  are as expected. The interesting result is that we also get
$$ \{\omega(x^{+}),\varphi(x^{+},x^{-}) \}^{*}=-{1\over4}
\int_{-L}^{L}dy^{-}
G^{Q}(x^{-},y^{-}){\lambda \varphi(x^{+},y^{-})-\mu \over  2 \lambda
\omega(x^{+}) - \mu} $$
This indicates that there is coupling between the  non-zero modes of the field
of
the scalar field and the background field $\omega$, the zero mode of the
scalar field.

There  is also coupling between the zero modes of the field and the momentum of
the
non-zero modes of the field :
$$ \{\omega(x^{+}),\pi_{\varphi}(x^{+},x^{-}) \}^{*}={1\over4}
\int_{-L}^{L}dy^{-}
G^{Q}(x^{-},y^{-}){\lambda \varphi(x^{+},y^{-})-\mu \over ( 2 \lambda
\omega(x^{+}) - \mu)^{2}} 2 \lambda $$
On the other hand we get the following Dirac brackets for the
zero mode of the field
$$ \{\omega(x^{+}),\pi_{\omega}(x^{+}) \}^{*}=
   \{\omega(x^{+}),\omega(x^{+}) \}^{*}=
   \{\pi_{\omega}(x^{+}),\pi_{\omega}(x^{+}) \}^{*}=0 $$
This means that upon Dirac quantization this quantity is not dynamical.
Nonetheless, since it depends on the quantized field $\phi$, and
since it couples to the non-zero part of the field, it is important
in studying non-trivial topological properties of the theory.
\chapter{ The c=2 Matrix Model  }
Let us study now this Dirac (constrained) quantization applied to
the $c=2$ matrix model introduced by Dalley and Klebanov \refmark\dk.
The Lagrangian is
$$\L = Tr\Big\{1/2(\partial_{\alpha}M)^{2}+1/2\mu M^{2}-{\lambda \over
3\sqrt{N} }M^{3} \Big\} $$
where $M(x^{-},x^{+})$ are $N\times N$ hermitian matrices.
We apply now the method of  constrained quantization described above
and get the following for the full Hamiltonian
$$P^{-}(x^{+}) = \int_{-L}^{L} dx^{-}Tr\Big\{{1\over2}\mu M^{2}-{\lambda
\over 3\sqrt{N} } M^{3} \Big\}+ $$
$$ +\int_{-L}^{L}dx^{-}Tr \Big\{U_{1}(x^{+},x^{-})\Theta_{1}(x^{+},x^{-})
\Big\} +Tr\Big\{\Theta_{2}(x^{+})U_{2}(x^{+})2L\Big\} $$
where the $U's$ are the new Lagrange multipliers and the $\Theta's$
the new constraints :
$$ \Theta_{1}(x^{+},x^{-}) =
\Pi_{\M}(x^{-},x^{+})-\partial_{-}\M(x^{-},x^{+}) \approx 0$$
and
$$\Theta_{2}(x^{+}) =\Pi_{\Omega}(x^{+}) \approx 0  $$

The analysis goes through as in the previous chapter - except that
now I have the extra indices, since $TrM^{2}$ means $\sum_{ij}M_{ij}
M_{ji}$; I'll suppress these indices from now on.
As before, I split the field $M$ into  a zero mode part
$$ \Omega(x^{+}) =  P*M(x^{+},x^{-}) =
{1\over 2L} \int_{-L}^{L} dx^{-} M(x^{+},x^{-})$$
and
$$ \M(x^{+},x^{-})=Q*M (x^{+},x^{-})=\delta *M(x^{+},x^{-})-
P*M(x^{+},x^{-}) =M(x^{+},x^{-})-\Omega(x^{+})$$
is the non-zero part. I get the following form for the new hamiltonian
$$\H_{P}=P^{-}(x^{+})=
 \int_{-L}^{L} dx^{-}Tr\Big\{{1\over2}\mu M^{2}-{\lambda
\over 3\sqrt{N} } M^{3} \Big\}+ $$
$$ +\int_{-L}^{L}dx^{-}Tr \Big\{
(\Pi_{\M}(x^{-},x^{+})-\partial_{-}\M(x^{-},x^{+}) )
U_{1}(x^{+},x^{-})\Big\} +
Tr\Big\{\Pi_{\Omega}(x^{+})U_{2}(x^{+})2L\Big\} $$
where the $\Pi_{\M}$ is the momentum canonical to $\M$ and $\Pi_{\Omega}$
the momentum canonical to $\Omega$.
The $U's$ are found to be
$$U_{1}(x^{+},x^{-}) = \int_{-L}^{+L}dy^{-} G^{Q}(x^{-},y^{-}){1\over4}
\Big\{\mu\M(x^{+},y^{-})+\mu\Omega(x^{+})+$$
$$+{\lambda \over 2!}(\M(x^{+},y^{-})^2
+ 2\M(x^{+},y^{-})\Omega(x^{+}) + \Omega(x^{+})^{2} ) \Big\} $$
and
$$U_{2}(x^{+})={-1 \over 2L}\int_{-L}^{L} dx^{-}
(2 \lambda \Omega(x^{+}) - \mu I)^{-1}
( \lambda \M(x^{-},x^{+}) - \mu I) U_{1}(x^{-},x^{+}) $$
As before,
the interesting part is that there is coupling between the zero mode
sector and the non-zero mode sector due to the following commutator
$$ \{\Omega(x^{+}),\M(x^{+},x^{-}) \}^{*}=-{1\over4}
\int_{-L}^{L}dy^{-}
(2 \lambda \Omega(x^{+}) - \mu I) ^{-1}
G^{Q}(x^{-},y^{-})(\lambda \M(x^{+},y^{-}) - \mu I) $$

In this case
there  is also coupling between the zero modes of the field and the momentum of
the
non-zero modes of the field :
$$ \{\Omega(x^{+}),\Pi_{\M}(x^{+},x^{-}) \}^{*}={1\over4}
\int_{-L}^{L}dy^{-}
( 2 \lambda \Omega(x^{+}) - \mu I)^{-2}
G^{Q}(x^{-},y^{-})
(\lambda \M(x^{+},y^{-})-\mu I) 2 \lambda $$
\chapter{Conclusions}
The constrained light-cone quantization indicates that there is
coupling between the zero mode sector and the nonzero mode sector.
This means that the analysis of Dalley and Klebanov \refmark\dk
needs to be redone in light of this result.  It is unclear at this point
if there is still the kind of excitations which Dalley and Klebanov
associated with Liouville mode.
This is because the type of critical behaviour studied might involve
excitations of the zero modes of the field which were previously left out. A
more
careful study is necessary to discover what happens now in the
 double scaling limit of this matrix  model.

There is also the possibility that the tachyonic mode which they discover
might be due to instabilities in the $\phi^{3}$ theory rather than
due to the bosonic string. In a recent paper, Hiller and Swenson
\REF\hs{ John Swenson  and John Hiller, 'Numerical signatures of
vacuum instability in a one-dimensional Wick Cutkosky model on the
light-cone,' Sept. 1992}
studied the Wick-Cutkosky model which is similar to the $\phi^{3}$
model considered by Dalley and Klebanov, and found instabilities in
vacuum , as expected for a cubic theory.
\chapter{Acknowledgements}
I would like to thank Prof. Stanley Brodsky for suggesting this problem
and to Prof. Blankenbecler for his continuing support .
I would also like to thank Igor Klebanov and Dave Robertson   for
useful discussions.
\endpage
\refout
\end
\REF\clcqschw{T. Heinzl, S. Krusche and E. Werner, 'The Fermionic
Schwinger Model in Light Cone  Quantization,' Phys. Lett. B, \us{275},
410, (1992) }
\refend
\REF\clcqschwa{T. Heinzl, S. Krusche and E. Werner, 'Non-Trivial Vacuum
Structure in Light Cone  Quantum Field Theory,' Nucl. Phys.  ,
 A, \us{532},429c, (1991) }
\refend
\REF\clcqschwb{T. Heinzl, S. Krusche and E. Werner, 'Non-Perturbative
Vacua in Light Cone  Quantum Field Threory,' Nucl. Phys. (Proc.
Suppl.), B, \us{23},182, (1991) }
\refend
\REF\clcqphi{T. Heinzl, S. Krusche and E. Werner,
'Light Cone  Quantization of Scalar Field Theories,' TPR-91-20 }
\refend
\REF\phitreeplus{T. Heinzl, S. Krusche and E. Werner, 'Non-Perturbative
 Light Cone  Quantum  Field Theories Beyond the Tree Level,' TPR-92-16 }
\refend
\REF\phitreezero{T. Heinzl, S. Krusche and E. Werner, 'Zero Mode
 Corrections in Light-Cone  Quantum  Field Theory ,' TPR-92-17 }
\refend
\REF\phitree{
T. Heinzl, S. Krusche and E. Werner, 'Spontaneous Symmetry Breaking
 in Light Cone  Quantum Field Theory' Phys. Lett. B, \us{272},
 54, (1991) }
\refend
\REF\ntrivac{
T. Heinzl, S. Krusche and E. Werner like in Phys. Lett. B,
\us{256},55 (1991)
or Z. Phys. A, \us{334}, 443 (1989); see also K. Hornbostel's preprint
CLNS 91/1078 (1991);Gary McCartor, Z. Phys. \us{C41}, 271 (1988)}
\refend
\REF\dirac{ P.A.M. Dirac, Rev. Mod. Phys.\us{21},392 (1949).}
\refend
\REF\alexmat{Matthias Burkardt and Alex Langnau, Phys. Rev.
\us{D44}, 3859 (1991)}
\refend
\REF\ohiost{Robert J. Perry, Avaroth Harindranath and Kenneth
G. Wilson, Phys. Rev. Lett. \us{65}, 2959 (1990); A. Harindranath and
Robert J. Perry, Phys. Rev. \us{D43}, 4921 (1991)}
\refend
\REF\bsdisc{Ovid C. Jacob, 'Discrete Symmetries for the Bound State
Problem of Positronium in Light-Cone Quantization,' SLAC-PUB-58xx,
 June 1992 .}
\refend
\REF\dba{P. A. M. Dirac, Can. Jour. Math., \us{1}, 1, (1950); 'Lectures
in Quantum Mechanics', Benjamin, NY, 1964;
P. G. Bergmann, Helv. Phys. Acta (Suppl.), \us{4}, 79 (1956) }
\refend .
\REF\zeropert{Gary McCartor and David Robertson, 'Bosonic Zero
Modes in DLCQ,' SMUHEP/91-04 Oct.   1991}
\refend
\end
\refout
\end